
\let\includefigures=\iftrue
%
\let\includefigures=\iffalse
%
\let\useblackboard=\iftrue
%
%
%
\input harvmac.tex
\message{If you do not have epsf.tex (to include figures),}
\message{change the option at the top of the tex file.}
\input epsf
\epsfverbosetrue
\def\fig#1#2{\topinsert\epsffile{#1}\noindent{#2}\endinsert}
\def\fig#1#2{}
%
\def\Title#1#2{\rightline{#1}
\ifx\answ\bigans\nopagenumbers\pageno0\vskip1in%
\baselineskip 15pt plus 1pt minus 1pt
\else
\def\listrefs{\footatend\vskip 1in\immediate\closeout\rfile\writestoppt
\baselineskip=14pt\centerline{{\bf References}}\bigskip{\frenchspacing%
\parindent=20pt\escapechar=` \input
refs.tmp\vfill\eject}\nonfrenchspacing}
\pageno1\vskip.8in\fi \centerline{\titlefont #2}\vskip .5in}

\ifx\answ\bigans\def\tcbreak#1{}\else\def\tcbreak#1{\cr&{#1}}\fi
\useblackboard
\message{If you do not have msbm (blackboard bold) fonts,}
\message{change the option at the top of the tex file.}
\font\blackboard=msbm10 scaled \magstep1
\font\blackboards=msbm7
\font\blackboardss=msbm5
\newfam\black
\textfont\black=\blackboard
\scriptfont\black=\blackboards
\scriptscriptfont\black=\blackboardss

\else

\fi
%
\def\yboxit#1#2{\vbox{\hrule height #1 \hbox{\vrule width #1
\vbox{#2}\vrule width #1 }\hrule height #1 }}
\def\fillbox#1{\hbox to #1{\vbox to #1{\vfil}\hfil}}
\def\ybox{{\lower 1.3pt \yboxit{0.4pt}{\fillbox{8pt}}\hskip-0.2pt}}
\def\comments#1{}

\def\p{\partial}

\def\tr{{\rm tr\ }}

\Title{\vbox{\baselineskip12pt
\hfill{\vbox{
\hbox{BROWN-HET-1020\hfil}
\hbox{hep-th/9510161}}}}}
{\vbox{\centerline{ }
\vskip20pt
\centerline{Boundary States of D-Branes and Dy-Strings}}}
\centerline{Miao Li}
\smallskip
\centerline{Department of Physics}
\centerline{Brown University}
\centerline{Providence, RI 02912}
\centerline{\tt li@het.brown.edu}
\bigskip
\noindent

Polchinski's recent construction of Dirichlet-branes of R-R charges, together
with Witten's mechanism for forming bound states of both NS-NS charges
and R-R charges, provides a rigorous  method to treat these dy-branes. We
construct the massless sector of boundary states of D-branes, as well as
of dy-strings of charge $(p,1)$. As a consequence, the string tension formula
predicted by duality in the type IIB theory is obtained.

\Date{October 1995}
\nref\ht{C. Hull and P. Townsend, Nucl. Phys. B438 (1995) 109,
 hep-th/9410167.}
\nref\witten{E. Witten, Nucl. Phys. B443 (1995) 85, hep-th/9503124.}
\nref\dghr{A. Dabholkar, G. Gibbons. J.~A.~Harvey and F.~Ruiz~Ruiz,
Nucl. Phys. B340 (1990) 33; A.~Dabholkar and J.~A.~Harvey, Phys.
Rev. Lett. 63 (1989) 478.}
\nref\johna{J. H. Schwarz, ``An SL(2,Z) multiplet of type IIB superstrings,''
CALT-68-2013, hep-th/9508143.}
\nref\dkl{M. J. Duff, R. R. Khuri and J. X. Lu, ``String solitons,''
hep-th/9412184.}
\nref\andy{A. Strominger, Nucl. Phys. B451 (1995) 96, hep-th/9504090.}
\nref\gms{B. R. Greene, D. R. Morrison and A. Strominger, Nucl. Phys.
B451 (1995) 109.}
\nref\bbs{K. Becker, M.~Becker and A.~Strominger, ``Fivebranes, membranes
and nonperturbative string theory,'' NSF-ITP-95-62, hep-th/9507158.}
\nref\joe{J. Polchinski, Phys. Rev. D50 (1994) 6041, hep-th/9407031.}
\nref\joep{J. Polchinski, ``Dirichlet-branes and Ramond-Ramond charges,''
hep-th/9510017.}
\nref\edw{E. Witten, ``Bound states of strings and p-branes,''
hep-th/9510135.}
\nref\hull{E. Bergshoeff, C. M. Hull and T.~Ortin, ``Duality in type II
superstring effective action,'' hep-th/9504081; C.~M.~Hull,
``String-string duality in ten dimensions,''
hep-th/9506160.}
\nref\johnb{J. H. Schwarz, ``The power of M theory,'' RU-95-68,
CALT-68-20254, hep-th/9510086.}
\nref\hs{G. Horowitz and A. Strominger, Nucl. Phys. B360 (1991) 197.}
\nref\dlp{J. Dai, R. G. Leigh and J. Polchinski, Mod. Phys. Lett. A4
(1989) 2073; R.~G.~Leigh, Mod. Phys. Lett A4 (1989) 2767.}
\nref\sen{A. Sen, Phys. Lett. B329 (1994) 217, hep-th/9402032.}
\nref\pc{J. Polchinski and Y. Cai, Nucl. Phys. B296 (1988) 91.}
\nref\clny{C. G. Callan, C. Lovelace, C. R. Nappi and S. A. Yost,
Nucl. Phys. B308 (1988) 221.}
\nref\ph{P. Horava, Nucl. Phys. B327 (1989) 461.}
\nref\green{M. B. Green, Phys. Lett. B329 (1994) 435.}
\nref\fms{D. Friedan, E. Martinec and S. Shenker, Nucl. Phys. B271
(1986) 93.}
\nref\edwi{E. Witten, Phys. Lett. B86 (1979) 283.}

\newsec{Introduction}
One of the most fruitful dualities in string theory is the string-string
duality in the type IIB string in ten dimensions \ht, it implies
U-duality in various dimensions below ten. Using U-duality,
Witten shows that the weakly coupled type II string, together with
weakly coupled eleven dimensional
supergravity, controls dynamics at strong coupling in various
regions of the moduli space of the type II strings \witten. The fundamental
string corresponds to a singular string solution to the low energy
effective action \dghr, carrying charge corresponding to the NS-NS second
rank antisymmetric field. Given the conjectured $SL(2,Z)$ duality, one
expects infinitely many string solitons carrying both NS-NS charges and R-R
charges. This is indeed true in the context of the low energy effective action,
solutions are recently constructed by Schwarz \johna, for pairs of coprime
integers $(p,q)$, where $p$ is the NS-NS charge and $q$ is the R-R charge.
In analogy with dyons, one might call these strings as dy-strings, since
the discrete duality group $SL(2,Z)$ acts on them exactly the same way
as it acts on dyons in some four dimensional quantum field theories.
In addition to dy-strings, there are solitonic solutions of p-branes
of NS-NS charges and R-R charges, for a review see \dkl. There are important
implications of the existence of such p-branes. Some applications can be
found for example in \andy\ \gms\ \bbs. Given the importance of these objects
in nonperturbative string theory,
one would like to ask whether it is possible to describe these p-branes
beyond the low energy limit. In a
couple of insightful papers, Polchinski points out that Dirichlet-branes
are nonperturbative objects and give rise to the characteristic nonperturbative
contribution of order $\exp(-O(1/g_s))$ \joe, and these branes indeed
carry R-R charges and the minimal Dirac quantization condition is
satisfied \joep. With this exact conformal field theory formulation at
hand, we take a minor step in this paper to construct the boundary
states for the D-branes, as well as for  $(p,1)$ strings based on a
construction of Witten \edw.

In the type IIB theory, in addition to the metric $G_{\mu\nu}$, antisymmetric
tensor field $B^1_{\mu\nu}$ and the dilaton $\phi$ in the NS-NS sector,
there are a second antisymmetric tensor field $ B^2_{\mu\nu}$ and a second
scalar $\chi$, and a self-dual fourth rank antisymmetric tensor field
$A_{\mu_1\dots\mu_4}$, all arise in the R-R sector. Einstein metric
$g_{\mu\nu}=G_{\mu\nu}e^{-\phi/2}$ is the natural metric to discuss
duality, since it is invariant under a duality transformation. Form
a complex field $\lambda=\chi+ie^{-\phi}$. Under the group element
$\Lambda=\left(\eqalign{a&\quad b\cr c&\quad d}\right)$, $\lambda$
transforms according to \hull
\eqn\dual{\lambda\rightarrow {a\lambda -b\over -c\lambda +d},}
and accordingly the doublet of antisymmetric tensor fields transforms
as $B\rightarrow \Lambda B$. The low energy effective action
remains invariant. The singular string solution of \dghr\ can be
used to generate a string soliton solution carrying charges $(p,q)$,
as in \johna. Existence of five-brane solutions carrying corresponding
``magnetic'' charges of the B fields implies that $p$ and $q$ are integers.
Furthermore, for a stable such solution, $p$ and $q$ must be coprime,
as the string tension formula dictates \johna. Let $\lambda_0$ be
the asymptotic value of the complex field $\lambda$, the $SL(2,Z)$ invariant
string tension is
\eqn\tension{T_{p,q}={1\over\sqrt{\hbox{Im}}\lambda_0}|p+q\lambda_0|T,}
as expected on a general ground, where $T$ is the fundamental string
tension. It should be mentioned that the above tension is measured
against the Einstein metric. In a very recent paper, Schwarz generalizes
this tension formula to include various p-branes \johnb.

Polchinski argues in \joep\ that one should identify the solitonic
brane solutions carrying R-R charge \hs\ with Dirichlet branes, as were
first introduced in \dlp. The following significant features of D-branes
support this identification. First, due to Dirichlet boundary conditions
obeyed by open strings moving along D-branes, the ground state of
such open string (explained as excitation of the D-brane) carries
R-R charge. Second, the amplitude between two parallel
D-branes separated by a transverse distance $y$ scales as $y^{p-7}$,
where $p$ is the dimension of D-branes. When $p=1$, we have strings,
and the amplitude behaves exactly as predicted by string solution
\dghr. Finally, the minimal Dirac quantization condition is satisfied
by the charge of the p-brane and the charge of its dual, the (6-p)-brane.
A general p-brane carrying both NS-NS charge as
well as R-R charge can be viewed as a bound state of p-branes of
NS-NS charge and D-branes. This is similar to bound states of electrons
and monopoles in $N=4$ supersymmetric Yang-Mills theory \sen.
This view indeed is proven in a very
recent paper of Witten \edw, where some intriguing mechanism is explained
for forming a bound state of charges $(p,q)$ with $p$ and $q$ coprime, for
strings and fivebranes. We shall refer to these branes as dy-branes.

That the fundamental string and the D-string can form bound states
is easy to see. The fundamental string carries a unit
of NS-NS charge. A macroscopic string solution shows that a test
string parallel to it does not feel any force, since the exchange
of graviton cancels exchange of $B^1_{\mu\nu}$ states. Imagine however
a test D-string parallel to this macroscopic string. Now since D-string
does not carry NS-NS charge, it does not feel the antisymmetric tensor
field. It still feels the gravitational force, and this force can be
shown to be attractive. (The solution of \dghr\ takes the
form $G_{00}=\exp(2\phi)$, where $G_{00}$ is the sigma-model
metric for the fundamental string, then $G'_{00}=\exp(\phi)$ is the
sigma model metric for the D-string. It is easy to see
that this gives rise to an attractive force on the D-string.)
This is the reason for forming bound states
of $n$ fundamental strings of the same orientation with a D-string. This
string carries charges $(p,1)$, therefore is expected to be stable.
Witten's recent paper \edw\ provides an exact treatment of such bound
states.

To study interaction of fundamental strings in the background of a dy-brane,
one convenient way is to construct a boundary state associated with this
dy-brane. The boundary state certainly encodes properties of the dy-brane.
We shall in sect.2 construct the massless sector of the boundary state
for a D-brane, after briefly reviewing Polchinski's construction.
Lacking a manifestly supersymmetric covariant formulation, a full state
is hard to construct. It may be useful to explore Green-Schwarz
light-cone formulation, although we shall not do it here.  We see directly
a R-R field strength
arise in the R-R sector of the boundary state. In sect.3, we  proceed to
employ Witten's construction of $(p,1)$ strings to compute the boundary
state. This state is manifestly supersymmetric, therefore confirms that
the string state is a BPS-saturated state. The string tension formula is
readily read off from the boundary state. The agreement of this formula with
the one predicted by duality shows that Witten's construction is correct.
Or turn it around, it provides another evidence for $SL(2,Z)$ duality
in the type IIB theory. The final section is devoted to a discussion.

\newsec{Boundary States of the D-branes}

First, let us explain the idea about D-branes of \joep. Consider
two D-branes, one locates at the transverse
position $X^i=0$, another at $X^i=y^i$, $i=p+1,\dots, 9$. The amplitude
between these two D-branes is caused by exchange of closed string
states. To compute it, one need to define a boundary state for each
D-brane, say $|B,0\rangle$ and $|B,y\rangle$. These boundary states
encode information as how closed string states are produced in the
presence of D-branes. The amplitude is then given by
\eqn\ampl{A(y)=\int_0^\infty
 dt\langle B,0|{1\over 4}(1+(-1)^F)(1+(-1)^{\tilde{F}})
e^{-t\Delta}|B,y\rangle,}
where the GSO projection is inserted, and $\Delta$ is the closed string
propagator. One alternative way to compute $A$ is to view the above
cylindric amplitude as the one-loop amplitude of the open string with
two ends attached to D-branes. The formula corresponding to this picture
is
\eqn\amplb{A(y)=V_{p+1}\int{d^{p+1}p\over (2\pi)^{p+1}}\int {dt\over t}
\sum e^{-t(p^2+m_i^2)},}
where the factor $V_{p+1}$ is the p-brane world-volume, its origin
in \ampl\ is the delta-function arising from the inner product which
enforces conservation of longitudinal momentum. The sum in the above
formula runs over all possible open string states, with all possible
periodic boundary conditions in the loop channel. Polchinski's result is
\eqn\repeat{\eqalign{A=&V_{p+1} \int {dt\over t}(2\pi t)^{-(p+1)/2}
e^{-ty^2/8\pi\alpha'^2}\cr
&\left[ {1\over 2q}\left(\prod_{n=1}^\infty({1+q^{2n-1}\over 1-q^{2n}})^8
- \prod_{n=1}^\infty ({1-q^{2n-1}\over 1-q^{2n}})^8\right)-
8\prod_{n=1}^\infty({1+q^{2n}\over 1-q^{2n}})^8\right],}}
where the first two terms come from the NS sector, and the third from
the Ramond sector. The sum is zero, reflecting the fact that with the presence
of the D-brane, some supersymmetry is unbroken. This point will be made
explicit later. Interpreted in the closed string tree channel,
the second term in \repeat\ comes from the R-R sector. Taking the infrared
limit in this channel, the contribution of the massless fields is picked
up \joep:
\eqn\repea{A=(1-1)V_{p+1}2\pi(4\pi^2\alpha')^{3-p}G_{9-p}(y),}
where $G_{9-p}(y)$ is the massless scalar Green function in the $9-p$
transverse dimensions. Let $T_p$ be the D-brane tension.
The force strength, due to the coupling of the D-brane world-volume to
the $p+1$ antisymmetric tensor, is read off from \repea,
\eqn\force{T^2_p=2\pi(4\pi^2\alpha')^{3-p}.}
The remarkable consequence of this formula is the minimal Dirac quantization
condition satisfied by the $p$-brane and its dual, the $(6-p)$-brane.

Our interest in this paper lies in constructing corresponding boundary
state for a D-brane as well as its generalizations.
For the usual open string with Neumann boundary conditions, boundary states
are discussed in \pc\ \clny\ \ph. For the D-instanton,
where  Dirichlet boundary condition is enforced in all coordinates,
the boundary state is constructed in \green. For our purpose, though,
a generalization of that construction is necessary for the following
reasons. First, we wish to see the explicit form of the anti-symmetric
tensor contained in the boundary state; second, supersymmetry is an
important ingredient in generalizing the D-branes to dy-branes. The second
reason requires a boundary state containing operators in all pictures.
An exact such boundary state including all oscillators is formidable.
Fortunately for our purpose, we need to know only the massless sector,
and a useful algebraic strategy for constructing this part is presented
in \clny.

Let us start with the D-instanton. (Throughout this paper we work in
Euclidean spacetime. Results for Minkowski spacetime are obtained
by Wick-rotation.) The boundary state $|B\rangle$
satisfies the boundary conditions \green,
\eqn\drcon{(\alpha^\mu_n-\tilde{\alpha}^\mu_{-n})|B\rangle = (\psi^\mu_m
+i\tilde{\psi}^\mu_{-m})|B\rangle=0,}
where the first equation corresponds to $\partial_\sigma X^\mu=0$, and
the index $m$ is a half-integer for the NS-NS sector and an integer for
the R-R sector. The boundary condition for the ghost and anti-ghost
remains the same as in the Neumann case. The boundary condition for
the super-ghosts conform with the boundary condition for the fermions.
$|B\rangle$ can carry a finite momentum due to Dirichlet boundary
condition, so one can convert it into a fixed point $X$ in spacetime
through Fourier transformation. We shall work with bosonized super-ghosts as
in \fms. The massless sector of the NS-NS part of $|B\rangle$ turns
out to be the same as in the Neumann case
\eqn\nsb{|B_0\rangle_{NS} =\sum_{s=-\infty}^\infty\left( V^\mu_s(p)
\tilde{V}^\mu_{-2-s}(p)-V^b_s(p)\tilde{V}^c_{-2-s}(p)+V^c_s(p)
\tilde{V}^b_{-2-s}(p)\right)
{1\over 2}(c_0+\tilde{c}_0)|\Omega\rangle,}
where the sum is over all pictures. We follow closely notations in
\clny. $|\Omega\rangle$ is the $SL(2,C)$ invariant state. For simplicity we
shall replace ${1\over 2}(c_0+\tilde{c}_0)|\Omega\rangle$ by
$|\Omega\rangle$. Operators $V_s$ in picture $s$ are obtained from the
operators in the base picture
\eqn\base{\eqalign{V_{-1}^\mu(p)&=\psi^\mu e^{-\phi}ce^{ipX},\quad
V^b_{-1}(p)=2\beta e^{-\phi}ce^{ipX},\cr
V^c_{-1}(p)&={1\over 2}\gamma e^{-\phi}ce^{ipX},}}
by applying the picture changing operator ${\cal X}$
\eqn\pchan{{\cal X}=ie^\phi\psi\cdot\p X+e^{2\phi}\p\eta b+2c\p\xi.}
The precise relation is $V_{s+1}={\cal X}V_s$. $V_s$ carries momentum $p$,
again for simplicity we will ignore index $p$ in all formulas below.
As the formula \nsb\ shows, the total super-ghost number is $-2$. It can
be checked that $|B_0\rangle_{NS}$ is BRST invariant,
and ${\cal X}|B_0\rangle_{NS}=\tilde{{\cal X}}|B_0\rangle_{NS}$.

In order to obtain the R-R part of $|B_0\rangle$, we demand half of
supersymmetry be unbroken when applied to the whole state.
To this end, we introduce supersymmetry generators \fms
\eqn\susy{Q^A_{-1/2}=\oint {dz\over 2\pi i}S^Ae^{-\phi/2}(z),}
and its counterpart in the right-moving sector. SUSY operators in
other pictures can be obtained by applying ${\cal X}$. The R-R part
of the boundary state takes the form
\eqn\rrbs{|B_0\rangle_R=\sum_s V_sL\tilde{V}_{-2-s}|\Omega\rangle,}
where $V_sL\tilde{V}_{-2-s}$ stands for
$$V^A_{s}L_{AB}\tilde{V}^B_{-2-s},$$
with
\eqn\rrop{V^A_{-1/2}=S^Ae^{-\phi/2}c e^{ipX},}
and its picture-changed forms.
Note that all operators carry momentum $p$. In determining $|B_0\rangle_R$
by requiring supersymmetry, one will need the following formulas
\eqn\commu{\eqalign{\{Q^A_r,V^\mu_s\}&=-{1\over 2}V^B_{r+s}
(p^\nu\gamma^\nu\gamma^\mu)_B^{\quad A},\cr
[Q^A_r,V^B_s]&=-{i\over\sqrt{2}}(C\gamma^\mu)^{AB}V^\mu_{r+s}.}}
Some irrelevant terms in the above (anti-)commutators are omitted,
as in \clny.

In the Neumann case, operators
must carry a zero momentum, as a consequence of the Neumann boundary
conditions. It turns out that in order to determine the R-R part, a
nonvanishing regulator momentum is needed \clny, and $p^\mu_L=-p^\mu_R$,
to be consistent with Neumann boundary conditions. As a consequence,
there is a solution to $(Q^A_r\pm \tilde{Q}^A_r)(|B_0\rangle_{NS}
+|B_0\rangle_R)=0$. However, in the Dirichlet case, momenta in both
sectors are the same, therefore there is a solution to
$(Q^A_r \pm i\tilde{Q}^A_r)(|B_0\rangle_{NS}+|B_0\rangle_R)=0$
\eqn\solu{|B_0\rangle_R=\pm{1\over\sqrt{2}}\sum_s V_sp_\mu\gamma^\mu
C\tilde{V}_{-2-s}|\Omega\rangle.}
One may regard the plus sign represents the D-instanton, and the
minus sign represents the anti-instanton. The reason for this is that
the Green function between two D-instantons is vanishing due to
cancelation between the NS-NS sector and the R-R sector, while the
Green function between one instanton and an anti-instanton is not
vanishing. Each term in \solu\ is of the form $p^\mu V\gamma^\mu C\tilde{V}$,
therefore can be interpreted as the derivative of a scalar field. This
scalar is just the  one arising from the R-R sector and appearing as
the real part in the complex scalar $\lambda$. Its derivative is
just its field strength. This shows that indeed the fundamental string
couples to R-R bosons through their field strength. We see that the
D-instanton carry a charge of $\chi$.

The above discussion is easily generalized to the Dirichlet p-brane.
Imagine that the world-volume of this p-brane coincide with
the subspace $X^\alpha$, $\alpha=0,\dots, p$. Its position can be
fixed at, say, $X^i=0$, $i=p+1,\dots, 9$, or it carries a finite momentum
$p^i$ in the transverse directions. The NS-NS part of the boundary operator
is simply
\eqn\nsba{|B_0\rangle_{NS} =\sum_s\left(- V^\alpha_s
\tilde{V}^\alpha_{-2-s}+V^i_s
\tilde{V}^i_{-2-s}-V^b_s\tilde{V}^c_{-2-s}+V^c_s
\tilde{V}^b_{-2-s}\right)|\Omega\rangle,}
where the sign in front of the longitudinal part is different
from that in front of the transverse part, since the boundary
condition along the world-volume is Neumann. Now as in the pure Neumann
case, one introduces nonvanishing $p^\alpha_L=-p^\alpha_R$ as
a regulator. The transverse components are the same in both the left-moving
and the right-moving sectors. After doing so, one can check that
the boundary state \nsba\ is BRST invariant. We shall use $p$  standing
for $(p^\alpha_L,p^i)$.

Again we impose half of supersymmetry to
obtain the R-R part. It is easy to see that both the combinations in
the pure Neumann case and the pure Dirichlet case do not work for
$p\ge 0$. An ansatz should be such that when $p=-1$, the combination
is the one for the D-instanton, and when $p=9$, it is the one in
the pure Neumann case. There are two choices
\eqn\susyg{Q^A_r\pm i\tilde{Q}^B_r(\gamma^0\dots\gamma^p)_B^{\quad A}.}
When $p=9$, $i\gamma^0\dots\gamma^9$ is just $\gamma^{11}$. Since
$\tilde{Q}^A$ is chiral, this matrix disappears in \susyg, and
one recovers  $Q^A_r\pm \tilde{Q}^A$, the combinations in the
pure Neumann case. It can be checked that the above combinations
are consistent with the mixed boundary conditions.
Requiring that the whole boundary state be
annihilated by $\susyg$, we are led to
\eqn\rrp{|B_0\rangle_R=\pm (-1)^{p(p-1)/2+1}{1\over\sqrt{2}}\sum_s
V_sp^\nu\gamma^\nu\gamma^0\dots\gamma^pC\tilde{V}_{-2-s}|\Omega
\rangle.}
Clearly, the massless R-R field is the field strength of the (p+1)-form
$A_{0\dots p}(X^i)$. This result agrees with that of Polchinski's which
is based on a general argument.
Again, the plus sign corresponds to the D-brane, and the minus
sign the Dirichlet anti-brane (with the opposite orientation of
its world-volume).

\newsec{$(p,1)$ Strings}

To describe a bound state of the D-string with charges $(0,1)$ and $p$
fundamental strings with charges $(1,0)$, it is helpful to compactify
one spatial dimension around which the D-string wraps once. Let
it be $X^1$. Now we follow Witten to construct a $(p,1)$ string.

Interaction of closed strings with the D-string is through open
strings whose both ends are attached to the D-string. The tree level
diagram is represented by a disk. The boundary of the disk,
attached to the world-sheet of the D-string, can wrap the circle
any number of times, it follows that the total winding number of
closed string states is not conserved. Consider the disk diagram
for $(1,0)$ strings with a nonvanishing antisymmetric
tensor $B^1_{\alpha\beta}$ (being in the NS-NS sector, it couples to
$(1,0)$ strings), this coupling part is
\eqn\wsa{S_B={1\over 2}\int d^2\sigma\epsilon^{ab}B^1_{\alpha\beta}
\p_aX^\alpha\p_bX^\beta.}
Under a gauge transformation $B^1_{\alpha\beta}\rightarrow B^1_{\alpha
\beta}+\p_\alpha\Lambda_\beta-\p_\alpha\Lambda_\beta$, the above action
is not invariant on a disk due to a boundary term. Such a term can be
absorbed in the coupling of the boundary to the U(1) gauge field
$A_\alpha$:
\eqn\uone{S_A=\int d\tau A_\alpha{dX^\alpha\over d\tau}.}
The combined action is invariant, provided $A_\alpha\rightarrow
A_\alpha+\Lambda_\alpha$ under the gauge transformation. Note
that $A_\alpha$ lives on the world-sheet of the D-string. There are also
scalar fields (with respect to the world-sheet of the D-string)
$A_i$, corresponding to the transverse displacement of the D-string
world-sheet, allowed by Dirichlet conditions \dlp. But these modes
are irrelevant for our discussion, because we are not considering
oscillating modes of the D-string. The gauge invariant field strength
of $A_\alpha$ is
$${\cal F}_{\alpha\beta}=F_{\alpha\beta}-B^1_{\alpha\beta},$$
with an action on the world-sheet of the D-string
\eqn\dws{{1\over 2\lambda}\int d^2X{\cal F}^2,}
where $\lambda$ is the string coupling constant, its appearance is
due to the fact that the above term arises from the disk amplitude.
Clearly this term provides a source for field $B^1_{\alpha\beta}$.
So if ${\cal F}\ne 0$, the D-string will carry a NS-NS charge.

A constant ${\cal F}$ is the only solution of the equation of motion,
which can be generated by placing a charge for $A_\alpha$ at infinity.
The conjugate momentum of $A$ is $\pi={\cal F}/\lambda$. If the abelian gauge
group is indeed $U(1)$, $\pi$ must be quantized, and so ${\cal F}_{\alpha
\beta}=p\lambda\epsilon_{\alpha\beta}$. This must be the case, for otherwise
annihilation of the D-string with its anti-string will result in
a continuous winding number state. Now, we are ready to construct the
boundary state for this $(p,1)$ string. Our problem is to impose certain
boundary condition for closed string which is emitted from the
$(p,1)$ string. This boundary condition is Neumann in longitudinal
directions, and Dirichlet in transverse directions. In addition to
this, the open strings move in the background of a constant field strength
on the world-sheet of the $(p,1)$ string. Luckily, this problem
for purely Neumann boundary condition is already addressed in
\clny.

Let us recall the result for a constant background ${\cal F}$ for purely
Neumann conditions. The NS-NS part of the boundary state is
\eqn\callan{|{\cal F}\rangle_{NS}=[\hbox{det}(1+{\cal F})]^{1/2}\sum_s
\{V^\mu_s\left({1-{\cal F}\over 1+{\cal F}}\right)_{\mu\nu}
\tilde{V}^\nu_{-2-s}+\hbox{ghost terms}\}|\Omega\rangle.}
The overall factor $[\hbox{det}(1+{\cal F})]^{1/2}$ is seen to
result from the path integral with a boundary term in the disk action.
The ghost terms are the same as in ${\cal F}=0$ case. The matrix
$(1-{\cal F})/(1+{\cal F})$ in \callan\ can be understood  as exercising
a rotation on the right-moving vector, and indeed is an orthogonal matrix.
Again in order to obtain the R-R part, one has to introduce momenta
with the constraint
\eqn\rela{p^\mu_L=-\left({1-{\cal F}\over 1+{\cal F}}\right)_{\mu\nu}
p^\nu_R.}
This is consistent with the rotation in \callan. Now it is a little
tricky to construct the surviving supersymmetry. Let it be
$$Q_{A,r}+M_A^{\quad B}\tilde{Q}_{B,r}.$$
It is found in \clny\ that the matrix $M$ can be expressed as
\eqn\mmt{M=[\hbox{det}(1+{\cal F})]^{-1/2}\exp(-{1\over 2}\gamma^\mu
\wedge\gamma^\nu{\cal F}_{\mu\nu}),}
where a peculiar convention is introduced: Upon expanding the exponential,
one assumes all gamma matrices are anti-commuting, therefore
there are only a finite number of terms. Now the solution to the
supersymmetry condition is unique
\eqn\call{|{\cal F}\rangle_R={i\over\sqrt{2}}\sum_s V_sp_L^\mu\gamma^\mu
\exp(-{1\over 2}\gamma^\mu
\wedge\gamma^\nu{\cal F}_{\mu\nu})C\tilde{V}_{-2-s}|\Omega\rangle.}

The obvious generalization of \callan\ to our case is
\eqn\neveu{|{\cal F}\rangle_{NS}=[\hbox{det}(1+{\cal F})]^{1/2}\sum_s
\{-V^\alpha_s\left({1-{\cal F}\over 1+{\cal F}}\right)_{\alpha\beta}
\tilde{V}^\beta_{-2-s}+V^i_s\tilde{V}^i_{-2-s}+\hbox{ghost terms}\}
|\Omega\rangle.}
The overall factor $[\hbox{det}(1+{\cal F})]^{1/2}$ is expected
by the path-integral argument. The above state is BRST invariant,
provided that the longitudinal momenta satisfy the same relation
as in \rela, and the transverse momenta in both sectors are the same.
Explicitly, we have
\eqn\data{\eqalign{\hbox{det}(1+{\cal F})&=1+p^2\lambda^2,\cr
{1-{\cal F}\over 1+{\cal F}}&={1\over 1+p^2\lambda^2}\left(\eqalign{1-
p^2\lambda^2
&\quad -2p\lambda\cr 2p\lambda &\quad 1-p^2\lambda^2}\right).}}
The matrix in \neveu\ has an off-diagonal term proportional to
$2p\lambda$, signaling an antisymmetric tensor $B^1_{\alpha\beta}$
as we expected.

We need to guess the correct combination of the supersymmetry.
This is easy, we need to recover supersymmetry for 1-brane discussed
in the previous section when ${\cal F}=0$, and to recover the purely
Neumann result when $p=9$.  There is a unique solution:
\eqn\susyd{Q_{A,r}+i\tilde{Q}^B_r\left(\gamma^0\gamma^1\exp({1\over 2}
\gamma^\alpha\wedge \gamma^\beta{\cal F}_{\alpha\beta})C\right)_{BA}
[\hbox{det}(1+{\cal F})]^{-1/2}.}
Requiring that the total boundary state be annihilated by the above
SUSY generator yields a unique solution
\eqn\rrr{|{\cal F}\rangle_R=-{1\over\sqrt{2}}\sum_s V_sp^\mu_L\gamma^\mu
\gamma^0\gamma^1\exp(-{1\over 2}
\gamma^\alpha\wedge \gamma^\beta{\cal F}_{\alpha\beta})C\tilde{V}_{-2-s}
|\Omega\rangle.}
Expanding the exponential in the above equation according to that
{\it ad hoc} rule, we see that there are two terms. One corresponds
to $B^2_{01}$, another corresponds to $\chi$. The latter is proportional
to $p\lambda$

The interaction strength between two $(p,1)$ strings can be read off
from the boundary state. The contribution from the NS-NS part determines
the string tension $T_{p,1}$
\eqn\sten{T_{p,1}^2={1\over 8}\left[\hbox{det}(1+{\cal F})
\left({1-{\cal F}\over 1+{\cal F}}\right)^2_{\alpha\beta}+8
\hbox{det}(1+{\cal F})-2 \hbox{det}(1+{\cal F})\right],}
where the first term comes from the longitudinal modes, the second
term comes from the transverse modes, and the third from ghosts.
An overall factor $1/8$ is used to normalize the result. Of course
a constant factor as a function of $\alpha'$ is omitted. A little calculation
then shows
\eqn\stens{T_{p,1}^2=1+p^2\lambda^2.}
This is exactly the result one expects from Duality \johna. It is also
interesting to compute the interaction strength coming from the
R-R sector. It is proportional to the square of the charge in terms of
the string's own antisymmetric tensor
\eqn\charge{\eqalign{\alpha_{p,1}^2&={1\over 8\times 2}\tr
\left({1\over 2}(1+\gamma^{11})\exp({1\over 2}\gamma^\alpha\wedge
\gamma^\beta{\cal F}_{\alpha\beta})
\exp(-{1\over 2}\gamma^\alpha\wedge \gamma^\beta{\cal F}_{\alpha\beta})
\right)\cr
&={1\over 32}\tr(1+\gamma^{11})(1+{1\over 2}\gamma^\alpha
\gamma^\beta{\cal F}_{\alpha\beta})(1-{1\over 2}\gamma^\alpha
\gamma^\beta{\cal F}_{\alpha\beta})
=1+p^2\lambda^2,}}
where the factor $1/8$ in the first equality is the same as in \sten,
an additional factor $1/2$ comes from the factor $1/\sqrt{2}$ in \rrr,
and the chirality projector is inserted because the $V^A$'s are chiral.
The fact that the above result is equal to $T_{p,1}^2$ should come
as no surprise, as expected for a BPS-saturated state.

Eq.\stens\ is the $(p,1)$ string tension measured in its own sigma model
metric, and in which the $(1,0)$ string tension is $\lambda$. When measured
in the $(1,0)$ string metric, one has
\eqn\reme{T^2_{p,1}={1\over\lambda^2}+p^2.}
In this metric, the $(1,0)$ string tension is $1$. One surprising point
about \reme, pointed out in \edw, is that the difference $T_{p,1}-T_{1,0}$
is of order $\lambda$. So the binding energy in the weak coupling limit
is almost equal to energy of all $(1,0)$ strings. The origin of this
property becomes
obvious by taking a look at the boundary state formula \neveu: That the
effect of gauge field $A_\alpha$ is nonlinear, makes it possible for
a D-string to carry a NS-NS charge with little cost of energy.

Finally, the integer charge $p$ can be shifted by an arbitrary amount $\chi_0$,
if one switches on a constant background of the R-R scalar $\chi$. This
effect is analogous to Witten's effect about the shift of electric charge
of a monopole when the theta-angle $\theta$ is switched on,  whose form is
$m+\theta/2\pi$ \edwi. With such a shift, our previous discussion
still goes through.

\newsec{Discussion}

The next logical step is to extend our discussion to include $(p,q)$
strings with $q>1$, and to include dy-branes. Witten in \edw\ suggests a
ansatz for such a string: That a $N=8$ $U(q)$ supersymmetric Yang-Mills
theory governs properties of $(p,q)$ strings. The gauge group
$U(q)$ is generated from Chan-Paton factors associated to $q$ D-strings.
To generate charge $p$ in the NS-NS sector, one places a quark
in the p-fold tensor of the fundamental representation at infinity.
The most intriguing of this scenario is the existence of adjoint
matter $X^i$, coming from transverse oscillations of the D-strings.
Its vacuum expectation value determines the separations of $q$
D-strings, yet they are noncommuting variables. A further understanding
of such system is necessary in order to extend our discussion
to construction of the boundary state of $(p,q)$ strings.
In the case of p-branes, we need understand more about the p-brane
solutions of NS-NS charges. Thus,
construction of the boundary states associated to dy-branes is
also for the future.

\noindent {\bf Acknowledgments}

We would like to thank A. Jevicki for helpful conversations, J. Polchinski
and J. Schwarz for correspondences. This work was supported
by DOE grant DE-FG02-91ER40688-Task A.

\listrefs

\end